\documentclass[aps,pra,twocolumn,showpacs,superscriptaddress]{revtex4-1}

\usepackage{amsfonts,mathrsfs,amsmath,amsthm,amssymb,bbold}
\usepackage{epsfig}
\usepackage{epstopdf}
\usepackage{txfonts}
\usepackage[dvipsnames]{xcolor}
\usepackage{multirow}
\usepackage{slashbox}

\begin{document}  
\title{Nuclear Magnetic Resonance model of an entangled sensor under noise}

\author{ Le Bin Ho}
\affiliation{Department of Physics, Kindai University, Higashi-Osaka, 577-8502, Japan}

\author{Yuichiro Matsuzaki}
\affiliation{Nanoelectronics Research Institute, 
National Institute of Advanced Industrial Science 
and Technology (AIST), 1-1-1
Umezono, Tsukuba, Ibaraki 305-8568, Japan}

\author{Masayuki Matsuzaki}
\affiliation{Department of Physics, Fukuoka University of Education, 
Munakata, Fukuoka 811-4192, Japan}

\author{ Yasushi Kondo}
\affiliation{Department of Physics, Kindai University, Higashi-Osaka, 577-8502, Japan}
\thanks{Electronic address: ykondo@kindai.ac.jp}


\begin{abstract}
Entangled sensors have been attracting much attention recently 
because they can achieve higher
sensitivity than those of classical sensors.
To exploit entanglement as a resource,
it is important to understand the effect of noise, because
the entangled state is highly sensitive to noise. 
In this paper, we present a Nuclear Magnetic Resonance 
(NMR) model of an entangled sensor in 
a controlled environment; 
one can implement the entangled sensor
under various noisy environments. 
In particular, we experimentally investigate the performance 
of the entangled sensor 
under time-inhomogeneous noisy environments, 
where the entangled 
sensor has the potential to surpass classical sensors.
Our ``entangled sensor'' consists of a multi-spin molecule
solved in isotropic liquid, and 
we can perform, or simulate, quantum sensing 
by using NMR techniques. 
\end{abstract}

\maketitle

\section{Introduction}

Quantum sensing has been attracting much attention recently
as an application of quantum mechanics
like quantum information technology \cite{Nielsen2000}, 
because it may achieve better sensitivity 
than a classical sensor. 
Quantum sensing may be divided into three categories according to what aspect of 
quantum mechanics employed to improve the performance 
of measurements: 
(i) quantum objects such as electrons or nuclear spins, 
(ii) quantum coherence such as superposition states or matter-wave-like nature, 
and 
(iii) quantum entanglement, which cannot be described classically
\cite{RevModPhys.89.035002}. 
The third may be the 
most quantum-like, and various efforts have been reported regarding this category.  
Among these, entanglement-enhanced magnetic field
sensing with atomic vapors was reported, such as 
spin squeezing (entanglement) within the atomic internal structure 
\cite{PhysRevLett.101.073601}, and 
employing entanglement between two vapor cells
\cite{PhysRevLett.104.133601}
to reduce noise. 
A more direct approach to enhance the sensitivity of measurements 
using entanglement was presented \cite{Giovannetti:2011aa},
and experimental efforts using trapped ions
\cite{Leibfried:2003aa,Leibfried1476,Leibfried:2005aa,PhysRevLett.86.5870}, 
ultra-cold atoms
\cite{Orzel2386,Appel10960}, and NMR
\cite{Jones1166,PhysRevA.82.022330}
have been reported. 

A potential problem involved in entangled sensors is their 
fragility to noise.
In fact, it has been theoretically proven
that an entangled sensor in a Markovian noisy environment,
where relaxation is exponential, cannot 
overcome the standard quantum limit (SQL) \cite{huelga1997improvement}. 
On the other hand, there are numerous theoretical 
predictions indicating that an entangled 
sensor can outperform a classical sensor 
under the effect of a time-inhomogeneous 
noisy environment, 
which induces a non-exponential decay 
\cite{PhysRevA.84.012103,chin2012quantum,tanaka2015proposed,RevModPhys.88.021002}.
However, there have been no experimental demonstrations yet for the latter case. 
Therefore, it is very important to study the behavior of entangled sensors 
under the effect of time-inhomogeneous noisy environments.

In this work, we investigate 
the behavior of an entangled sensor in 
a controlled environment
using NMR. 
Although the demonstration 
does not constitute a formal proof of the quantum-mechanical enhancement of an entangled
sensor in a real environment,
our achievements provide important information toward 
understanding the properties of an entangled sensor:
(i) As the performance of the entangled sensor strongly 
depends on the properties of the environment, 
systematic experimental analysis of 
the entangled sensor operating under various types of noise
is essential for the realization of quantum-enhanced sensing,
and therefore our experimental investigation with the NMR model
provides insights to assess the practicality of quantum sensors.
(ii)
Our NMR method can be implemented 
with commonly available experimental apparatus 
(which almost every university owns) 
at room temperature, and therefore experimentalists 
can use this model as a testbed or a simulator 
before attempting to construct a real entangled sensor. 
For example, experimentalists can first use our NMR model to evaluate  
the pulse sequences 
that are expected to apply to 
the entangled sensor in a real environment.

The remainder of the paper is organized as follows. In Section~\ref{sec2}, we closely follow 
References~\cite{Jones1166,PhysRevA.82.022330} and review 
how an entangled sensor is simulated using a star topology molecule.
We then present a method demonstrating
how to prepare a
controlled environment
following Ref.~\cite{Binho2019}.
Note that an engineered noisy environment 
in Ref.~\cite{Binho2019} is equivalent to a controlled one 
in this paper.
Finally, these two ideas are combined, and 
we simulate the entangled sensor 
in a controlled environment.
We present experimental results in Section~\ref{sec3}, 
where the dynamics of 
the entangled sensor are evaluated in a  time-inhomogeneous noisy environment, and a
successful application of a dynamical decoupling 
technique~\cite{RevModPhys.88.041001}
applied to the entangled sensor  is demonstrated.  Finally, Section~\ref{sec4} presents 
a summary of our findings.

\section{Theory}\label{sec2}

In this section, we describe our strategy to combine two ideas: 
(i) use of a star-topology molecule as an entangled magnetic sensor, and 
(ii) controlling the
environment that surrounds the sensor.

\subsection{Molecules as a Simulator of an Entangled Magnetic Sensor}
\label{TH_M_S}
Assume an isolated nuclear spin with state 
\begin{eqnarray}
|+\rangle =  \frac{|0\rangle + | 1\rangle}{\sqrt{2}},
\end{eqnarray}
where $|0\rangle$ and $|1\rangle$ 
are two eigenstates of the standard Pauli matrix $\sigma_z$.
The system is exposed to a magnetic field 
$B\, \vec{z}$, where $\vec{z}$ is the unit vector along the $z$-axis,
for a period $\tau$, and becomes 
\begin{eqnarray}
|+_\tau \rangle =  \frac{|0\rangle + e^{i \gamma_{\rm G} B \,\tau}
| 1\rangle}{\sqrt{2}},
\end{eqnarray}
where $\gamma_{\rm G}$ is the gyromagnetic ratio. 
Therefore, the acquired phase $\gamma_{\rm G} B \, \tau$ 
can be used to measure $B$. 
The sensitivity of a set of $N$ spins 
is proportional to $\sqrt{N}$,
which is the SQL~\cite{PhysRevLett.79.3865,Giovannetti:2011aa}. 

Now, if we assume that our sensor 
consists of $N$ spins and that
the initial state is entangled, such that
\begin{eqnarray}
|+_{\rm ent}\rangle =  \frac{|00 \dots 0\rangle 
+ | 11 \dots 1\rangle}{\sqrt{2}},
\end{eqnarray}
then, this state will evolve to  
\begin{eqnarray}
|+_{\rm ent, \tau}\rangle =  \frac{|00 \dots 0\rangle 
+ e^{i N \gamma_{\rm G} B \tau}| 11 
\dots 1\rangle}{\sqrt{2}},
\end{eqnarray}
after time $\tau$ elapses, and thus the sensitivity is proportional to 
the number of spins $N$, 
which is the Heisenberg Limit~\cite{PhysRevLett.79.3865,Giovannetti:2011aa}.
 
\begin{figure}[t]
\begin{center}
\includegraphics[width=8.6cm]{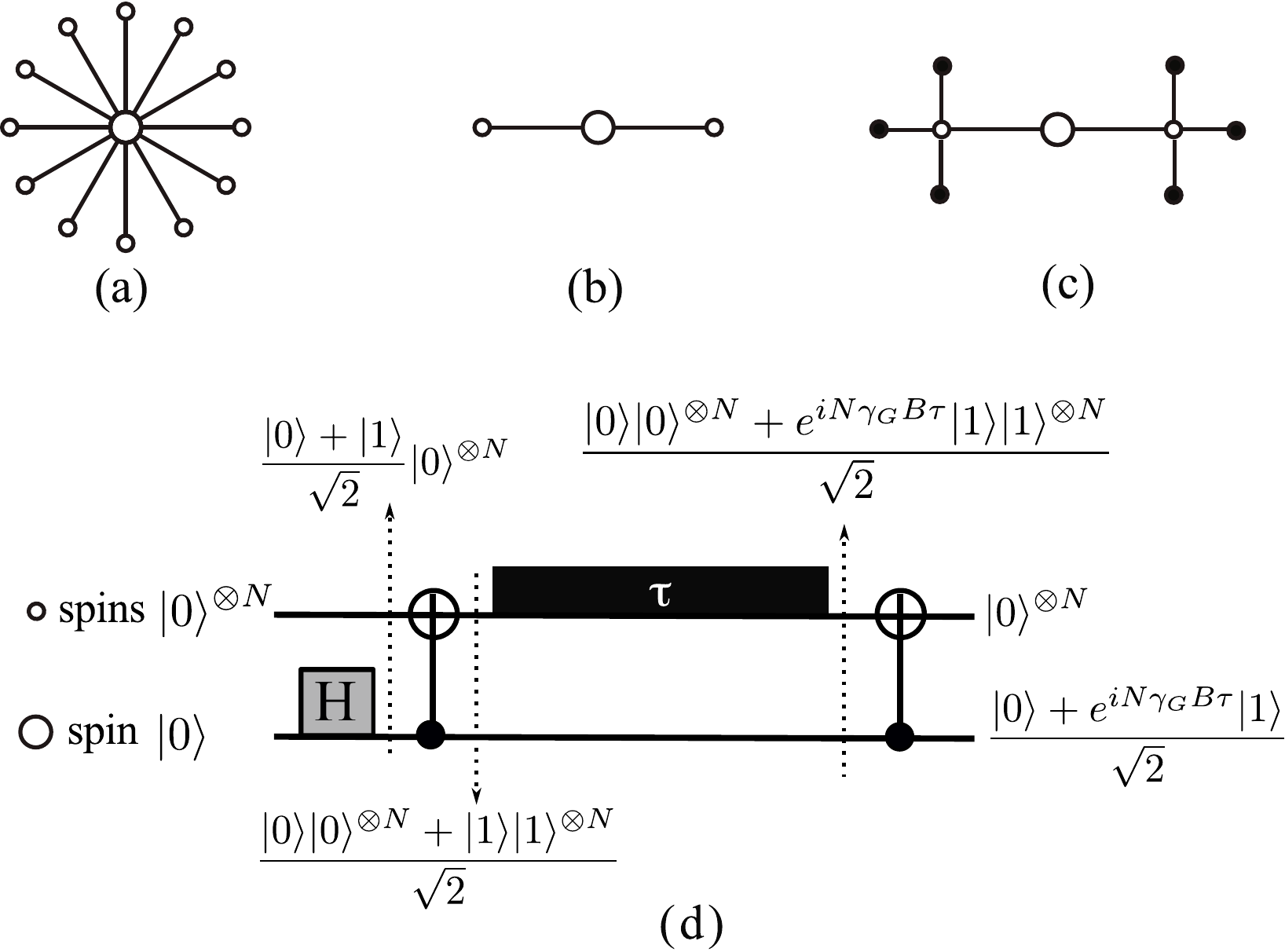}
\caption{
(a, b, and c) Three different interaction topologies among spins 
($\bigcirc, \circ$, and $\bullet$) discussed in this study 
and (d) quantum circuit for an entangled magnetic field 
sensor simulation. 
(a) Sketch of a star-like interaction structure among spins, and 
(b) the simplest star topology structure.
The small open circles ($\circ$'s) 
in (a) and (b) play the role of 
entangled sensors. 
(c) Two-step star topology structure used to conduct an entangled magnetic 
sensor simulation under a
controlled environment.
The large open circle ($\bigcirc$) is called the center spin, 
while the small open circles ($\circ$'s) 
are the side spins and play the role of 
an entangled sensor.
The six surrounding solid circles ($\bullet$'s) 
are introduced in order to generate 
a time-inhomogeneous noisy environment
acting on the entangled sensor. 
We refer to them as the environmental spins. 
(d) Basic quantum circuit used for measurements. 
Because of the symmetry of the interaction
structure,
all $\circ$ spins 
can be accessed globally. 
}
\label{fig:q_circuit}
\end{center}
\end{figure}

Jones et al.\ demonstrated the above measurement scheme with 
star-topology molecules, as 
schematically shown in Fig.~\ref{fig:q_circuit}(a).
They employed two molecules, 
trimethyl phosphite (TMP) \cite{Jones1166} and 
tetramethylsilane (TMS) \cite{PhysRevA.82.022330}.  
A TMP (or TMS) molecule consists of 
a center $^{31}$P  ($^{29}$Si) 
and three (four) methyl groups, and 
thus the center $^{31}$P  ($^{29}$Si)  
is surrounded by 9 (12) $^1$H spins. 
The highly symmetric structures of 
these molecules allow addressing of
all surrounding spins 
(small open circles in Fig.~\ref{fig:q_circuit}(a))
globally, and thus the operations required to measure 
a magnetic field can be simplified. 

In this work, we employed the simplest star topology
molecule, which consists of a center spin ($\bigcirc$) and
two side spins ($\circ$'s), 
as shown in Fig.~\ref{fig:q_circuit}(b).

We take the initial density matrix 
$\displaystyle |0\rangle \langle 0 |
\otimes \frac{\sigma_0}{2}  \otimes \frac{\sigma_0}{2} $
\cite{kondo2016} (see also the Appendix). 
The state of the center spin is $|0\rangle \langle 0 |$, 
while the two side spins  
are in a fully mixed state. 
When a magnetic field is applied only to the center spin ($\bigcirc$),
i.e., the star-topology structure is not effective, 
the density matrix becomes
\begin{eqnarray}
\rho_\bigcirc &=& 
 \frac{1}{8}
{\small \left( 
\begin{array}{cc}
 1 & e^{-i\theta}\\
e^{i\theta} & 1
\end{array} 
\right) }
\otimes \sigma_0 \otimes \sigma_0,
\end{eqnarray}
after the magnetic field is applied.
Here, $\theta = \gamma_{\rm G}B \tau$, 
and the subscript $\bigcirc$ stands for 
the case when the center spin is exposed to the field.
This case corresponds to 
non-entangled sensor.

Next, consider the case 
when the magnetic field is applied to the side spins ($\circ$'s),
as shown in Fig.~\ref{fig:q_circuit}(d). 
This case corresponds to an entangled sensor.
In a frame that co-rotates with the center spin,
it does not acquire 
a phase during the period $\tau$.
The initial state 
$\displaystyle \frac{1}{4} | 0 \rangle \langle 0 | \otimes 
\sigma_0 \otimes \sigma_0$
can be decomposed to
\begin{eqnarray}
 \frac{1}{4} \Bigl( 
  |000\rangle \langle 000| 
+ |011\rangle \langle 011| 
+ 
|001\rangle \langle 001| + |010\rangle \langle 010| 
\Bigr),
\end{eqnarray}
and thus,  
the final density matrix after the measurement operation is 
\begin{eqnarray}
\label{eq_meas_dm}
\rho_\circ
&=&  
{\small 
\frac{1}{8} \left(
\begin{array}{cccccccc}
 1 & 0 & 0 & 0 & e^{-2i\theta} & 0 & 0 & 0 \\
 0 & 1 & 0 & 0 & 0 & 1 & 0 & 0  \\
 0 & 0 & 1 & 0 & 0 & 0 & 1 & 0  \\
 0 & 0 & 0 & 1 & 0 & 0 & 0 & e^{2i\theta} \\
 e^{2i\theta} & 0 & 0 & 0 & 1 & 0 & 0 & 0 \\
 0 & 1 & 0 & 0 & 0 & 1 & 0 & 0 \\
 0 & 0 & 1 & 0 & 0 & 0 & 1 & 0 \\
 0 & 0 & 0 & e^{-2i\theta} & 0 & 0 & 0 & 1 \\
\end{array} \right) }\nonumber \\
&=& \frac{1}{8}
{\small \left( 
\begin{array}{cc}
 1 & e^{-2i\theta}\\
e^{i2\theta} & 1
\end{array} 
\right) }
\otimes 
|00\rangle \langle 00| \nonumber  \\
&+&\frac{1}{8}
{\small \left( 
\begin{array}{cc}
 1 & e^{2i\theta}\\
e^{-i2\theta} & 1
\end{array} 
\right) }
\otimes 
|11\rangle \langle 11| \nonumber \\
&+& \frac{1}{8}
\left( {\small \begin{array}{cc}
 1 & 1\\
 1 & 1
\end{array}} \right) \otimes 
\Bigl(|01\rangle \langle 01| + |10\rangle \langle 10| \Bigr),
\end{eqnarray}
where $\circ$ stands for the case when 
the magnetic field is applied to the side spins.
Note that owing to the twice as large
phase accumulation ($2\theta$, instead of $\theta$) of the 
initial states of $|000\rangle \langle 000|$ and 
$|011\rangle \langle 011|$,
the sensitivity of the side spins (the entangled sensor)
interacting with the magnetic fields
becomes twice as large as that of the center spin
(non-entangled sensor). 
 
\begin{figure}[t]
\begin{center}
\includegraphics[width=8.6cm]{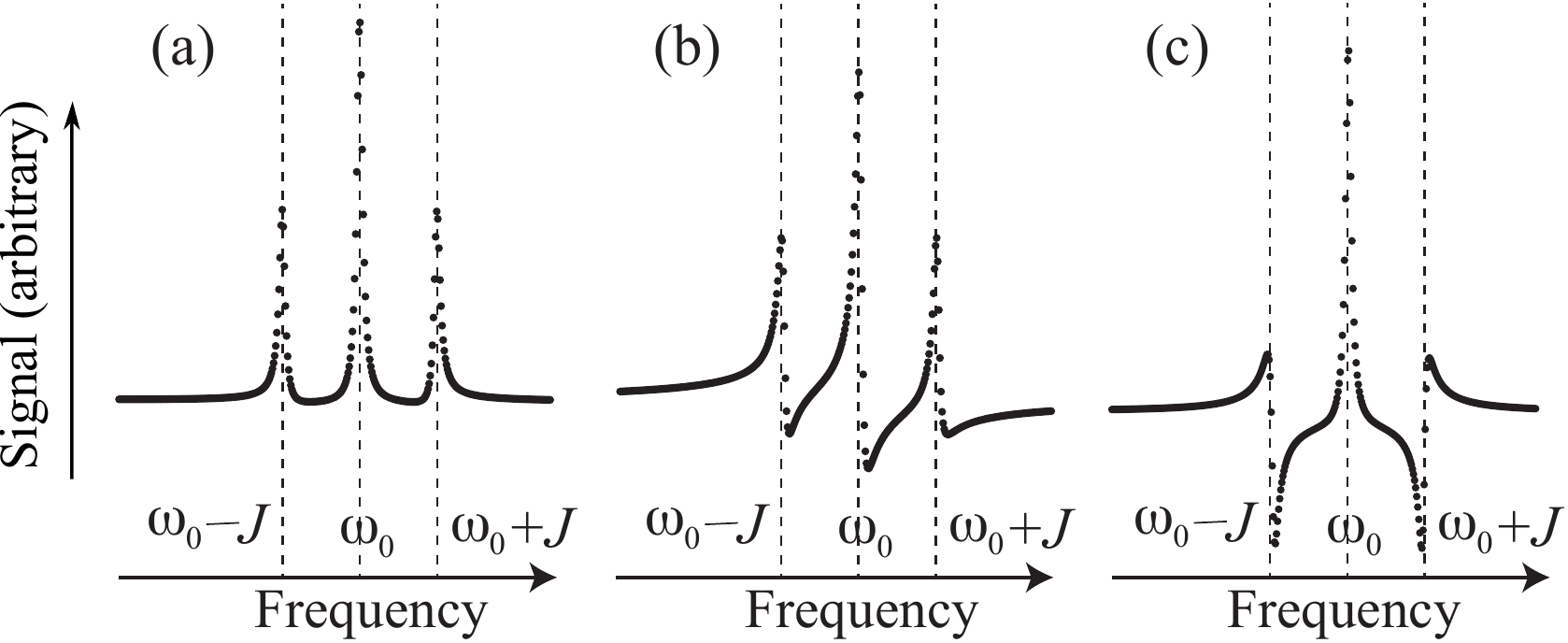}
\caption{
Theoretical spectra calculated from $S_k(t)$ 
for the case when $J T_2 = 22$. 
(a) $\theta =  0^\circ$, 
(b) $\theta = 50^\circ$ for $S_{\bigcirc}(t)$, 
(c) $\theta = 50^\circ$ for $S_{\circ}(t)$.
The frequency difference of these peaks is $J$, 
and the frequency of the center peaks is $\omega_0$.
}
\label{fig_fid_th}
\end{center}
\end{figure}

Next, the method of how to detect the acquired phases in the NMR is presented. 
The state $\rho_k$ ($k = \bigcirc, \circ$) is assumed to develop 
under the Hamiltonian
\begin{eqnarray}
 H&=& \omega_0 \frac{\sigma_z}{2} \otimes \sigma_0 \otimes \sigma_0\nonumber\\
&+&
J \left(\frac{\sigma_z}{2} \otimes \frac{\sigma_z}{2} \otimes \sigma_0 
+  \frac{\sigma_z}{2} \otimes \sigma_0 \otimes \frac{\sigma_z}{2} \right), 
\end{eqnarray}
where $\omega_0$ corresponds to a Larmor frequency of the center 
spin and $J$ is a coupling constant between the center spin and 
the two side spins. 
$\omega_0$ is introduced to identify the signal originating from 
the center spin in the frequency domain signal.
The signal from the center spin is 
\cite{Cory1634,Gershenfeld350,liquidNMRQC}
\begin{eqnarray}
 S_k(t) &=& {\rm Tr}\left( 
\left((\sigma_x+i \sigma_y) \otimes \sigma_0 \otimes \sigma_0 \right) 
e^{-i H t}\rho_ke^{iHt}
\right).
\end{eqnarray}
Therefore, the expected normalized signals are 
\begin{eqnarray}
 S_\bigcirc(t) 
 &=& \frac{1}{4}e^{-t/T_2}
\left( e^{-iJt} + 2 +  e^{iJt} \right)\cos(\omega_0 t + \theta), \label{Sl}\\
 S_\circ(t) 
&=&  \frac{1}{4}e^{-t/T_2}
\left( e^{-iJt-i2\theta} + 2 +  e^{iJt+i2\theta} \right)\cos \omega_0 t \label{Ss}, 
\end{eqnarray}
where $T_2$ is a phenomenological transverse relaxation time constant 
that is introduced 
for the signal to decrease exponentially. 
Equation~(\ref{Sl}) corresponds to the signal when the field is applied to the 
center spin (non-entangled sensor), 
while Eq.~(\ref{Ss}) corresponds to the case 
when the field is applied to the side spins (entangled sensor). 
Note the difference in the position of $\theta$ in $S_k(t)$.  
The three terms in parentheses in $S_k(t)$ correspond to three
peaks that are observed when the $S_k(t)$'s are Fourier transformed 
(see Fig.~\ref{fig_fid_th}).

\subsection{Controlled environment}
\label{TH_E_E}
\begin{figure}[t]
\begin{center}
\includegraphics[width=6.cm]{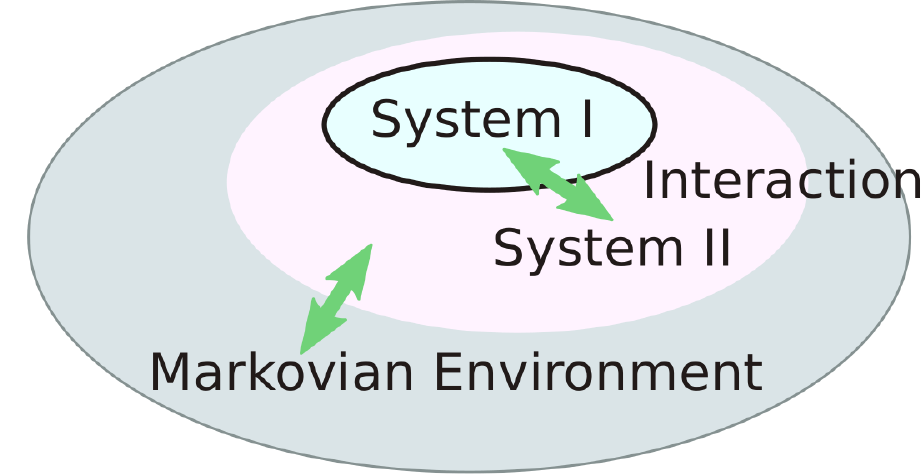}
\caption{(Color online) 
Controlled environment.
Herein, the Markovian environment
indirectly interacts with System~I through System~II.
}
\label{fig:model}
\end{center}
\end{figure}

Our idea to control the environment 
is demonstrated in Fig.~\ref{fig:model}.
If System~I directly interacts with a Markovian environment,
it decays exponentially. If it interacts with a
Markovian environment through System~II, it exhibits various 
decay behaviors, because System~II acts as a memory of the 
controlled non-Markovian environment
formed by System~II and the Markovian environment
\cite{kondo2016,Iwakura2017,Kondo2018,kondo2007,PhysRevA.55.2290,PhysRevLett.120.030402}. 
  
The chain of spins ($\circ-\bigcirc-\circ$ in Fig.~\ref{fig:q_circuit}(c)) 
is regarded as the sensor where the side spins (two $\circ$'s) 
accumulate the phase 
under the external field, and this phase is measured via the center spin
($\bigcirc$), as discussed in Sect.~\ref{TH_M_S}. The side spins ($\circ$'s)
are surrounded 
by two sets of three spins (three $\bullet$'s), which we call 
environmental spins. 
We regard the side spins (the two $\circ$'s) as two independent System~I's, while we consider 
the two sets of $\bullet$'s as two System~II's. These environmental 
spins, with the Markovian environment outside them, act 
as a time-inhomogeneous noisy environment 
applied to the side spins (two $\circ$'s) as we previously discussed in 
Refs.~\cite{kondo2016,Iwakura2017,Kondo2018,kondo2007,PhysRevA.55.2290,PhysRevLett.120.030402}. 
Note that only the nearest neighbor 
interactions are assumed important in this discussion.
After all, we can control
the environment of the sensor 
($\circ-\bigcirc-\circ$). 

\section{Experiments}\label{sec3}
In this section, we describe our approach 
to simulate an entangled sensor in a 
controlled environment
with a molecule solved 
in an isotropic liquid. First, we show how to 
prepare 
the Markovian controlled environment
and then discuss how to simulate the entangled sensor in
the controlled environment.

\subsection{
Markovian controlled environment}

Solute molecules in an isotropic liquid 
move rapidly and are influenced by 
a strong external magnetic field.
Thus, the interactions among nuclear spins in the solute 
and solvent molecules are averaged out~\cite{Levitt2008}.
In other words, the solute molecules are approximated 
to be isolated systems.

To control a Markovian environment,  
magnetic impurities, such as Fe(III),
are added to the solution.
Because the added magnetic impurities move rapidly and randomly,
they produce a Markovian environment, which
flip-flops the solute molecules' nuclear spins randomly.
The flip-flopping rate is proportional to the concentration 
of the magnetic impurities~\cite{Iwakura2017}. 
Moreover, we emphasize that
the nuclear spins of System~II 
are more strongly influenced by
the magnetic impurities than the inner ones (System~I)
because the dipole-dipole interactions
are short-range~\cite{Binho2019}.

\subsection{Sample and controlled environment}
\begin{figure}[t]
\begin{center}
\includegraphics[width=8.6cm]{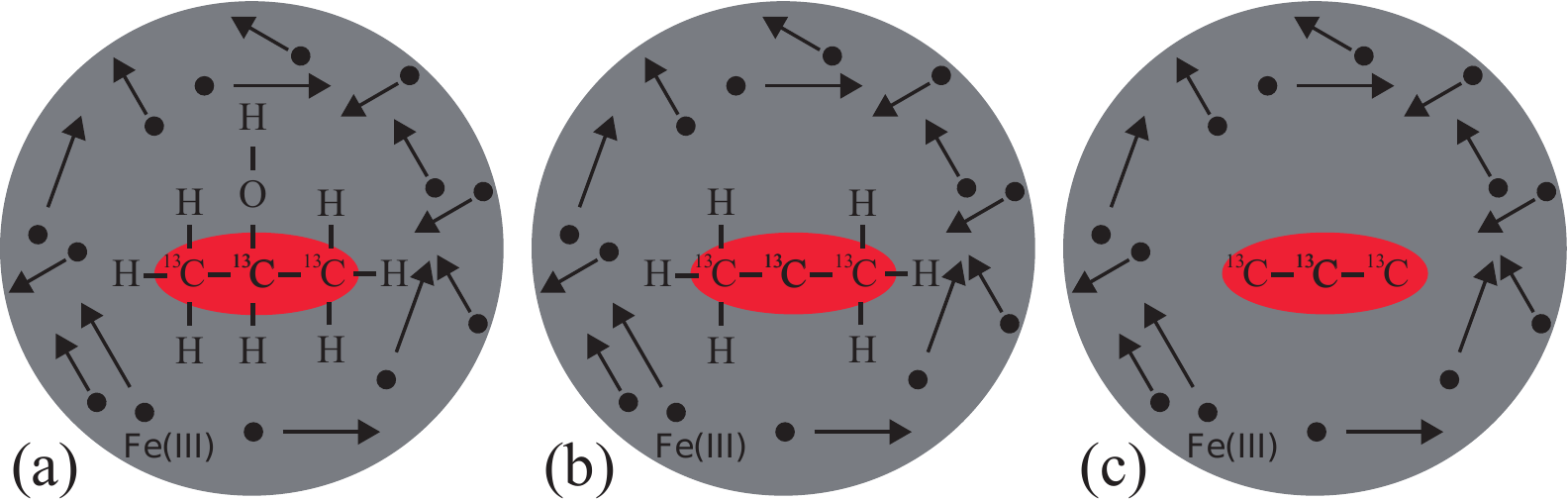}
\caption{(Color online) 
Sketch of a 2-propanol molecule in acetone-d6 with added magnetic impurities
(Fe(III) ions). 
The chain of the $^{13}$C spins is surrounded by 
the H spins. These H spins can
be selectively 
nullified by decoupling techniques, which provide
three different controlled environments.
The gray areas correspond to the System~II's and 
Markovian environments. 
We consider three cases:
(a) without decoupling, (b) selective decoupling of the H spin
attached to the center $^{13}$C, 
and (c) full decoupling of all H spins.
}
\label{fig_st}
\end{center}
\end{figure}

\begin{figure}[t]
\begin{center}
\includegraphics[width=8.6cm]{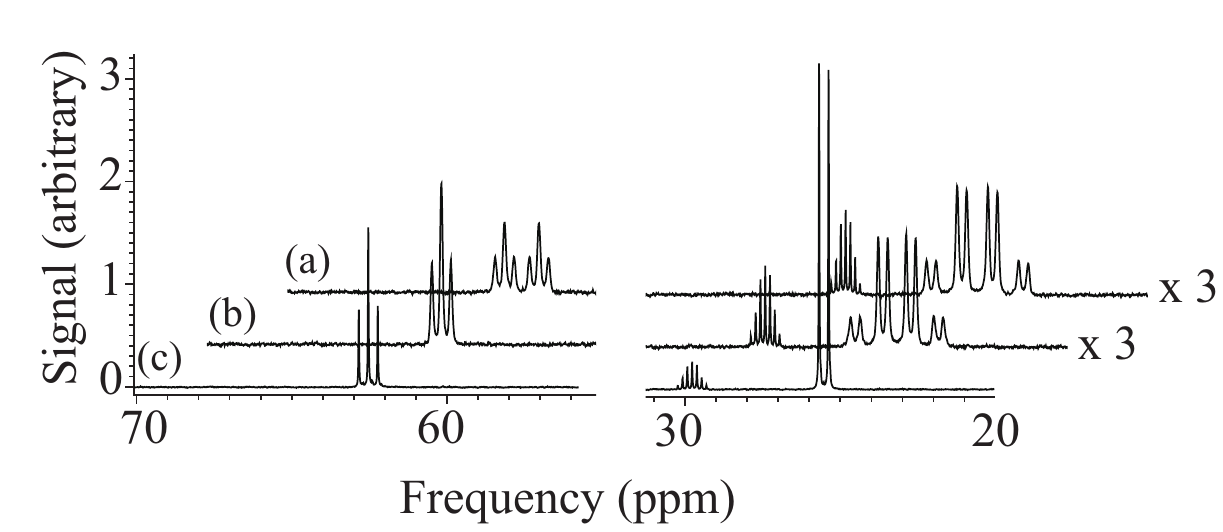}
\caption{
Spectra of the $^{13}$C spins in different decoupling conditions: 
(a) without decoupling, (b) selective decoupling by applying a small
continuous RF excitation, whose frequency is the Larmor frequency of HC (the H spin attached to CC), and 
(c) full decoupling by applying the WALTZ sequence to all H spins.
The sample is 0.41~M ${}^{13}$C-labeled 2-propanol in acetone-d6 with 12~mM of 
Fe(III)acac. 
The peaks at 62.6~ppm are originated from the center $^{13}$C of the chain,
while those at 25.5~ppm are originated from the side $^{13}$C's. 
The spectra in (a) and (b) are 
magnified three times.
}
\label{fig_sp}
\end{center}
\end{figure}

The two-step star-topology molecule that we employ in this work is 
2-propanol solved in acetone d6 with Fe(III) magnetic impurities added. 
The structure of this molecule is shown in Fig.~\ref{fig_st}(a). 
The three $^{13}$C spins correspond to
$\circ-\bigcirc-\circ$ in Fig.~\ref{fig:q_circuit}(b, c), 
while the H spins are employed as System~II in Fig.~\ref{fig:model}. 
We can selectively nullify the H spins by dynamical 
decoupling techniques~\cite{Levitt2008,RevModPhys.88.041001},
as shown in Fig.~\ref{fig_st}(a, b, c): 
(a) without decoupling, (b) selective decoupling of the H spin
attached to the center $^{13}$C 
(hereafter, referred to as the selective-decoupling case), 
and (c) full decoupling of all H spins
(hereafter, referred to as the full-decoupling case). 
This implies that 
the behavior of an entangled magnetic sensor in 
three different 
controlled environments
can be studied. 
Note that the interaction topology of Fig.~\ref{fig:q_circuit}(c) 
can be realized in the case of Fig.~\ref{fig_st}(b). The 
obtained spectra shown in Fig.~\ref{fig_sp} 
exhibit clear differences according to the interaction topology
differences presented in Fig.~\ref{fig_st}(a, b, c).

The Hamiltonian governing the nuclear spin dynamics of 2-propanol 
is given as
\begin{eqnarray}
\label{eq_mol_ham}
 {\mathcal H} &=& 
\sum_{j} \omega_0^{(j)} \frac{\sigma_z^{(j)}}{2}
 + \sum_{j<k} J^{(j,k)} \frac{\sigma_z^{(j)}\otimes\sigma_z^{(k)}}{4},
\end{eqnarray}
because $|\omega_{0}^{(j)} -\omega_{0}^{(k)}| \gg J^{(j,k)}$;
$\omega_{0}^{(j)}$ is the isotropic chemically shifted Larmor 
frequency of 
the $j$'th spin, and $J^{(j,k)}$ represents the interaction strength between 
the $j$'th and $k$'th spins~\cite{Levitt2008,Kondo2018}. 
$\omega_{0}^{(j)}$ and $J^{(j,k)}$ were measured from the spectra 
of a sample without magnetic impurities 
and are summarized in Table~\ref{H_parameters}.

\begin{table}[h]
\caption{\label{H_parameters}
$\omega_{0}^{(j)}$ and $J^{(j,k)}$ are summarized. 
We label the spins as 
CC (the center $^{13}$C in the C spin chain), 
CSs (the two side $^{13}$C's in the C spin chain), 
HC (the H spin attached to CC) and 
HSs (the H spins attached to a CS). 
Diagonal elements are the $\omega_{0}^{(j)}$'s 
in ppm, while the off-diagonal elements are those of $J^{(j,k)}$ in 
rad/s. NR implies that these values are too
low to be measured reliably. 
}
\begin{center}
\small\addtolength{\tabcolsep}{0.pt}
\begin{tabular}[t]{|c|c|c|c|c|}
\hline
 \backslashbox{$j$}{$k$}  
&CC & CSs  & HC & HSs \\
\hline 
CC &62.6 ppm & $ 
2\pi \cdot 38.4$ rad/s & $2\pi\cdot 140$ rad/s
& $2\pi\cdot 4.4$ rad/s \\
\hline 
CSs & & 25.5 ppm & NR &$2\pi \cdot124$ rad/s\\
\hline
HC & & & 3.78 ppm & NR \\
\hline
HSs  &&&& 1.21 ppm \\
\hline 
\end{tabular}
\end{center}
\end{table}

\subsection{Simulation of Entangled Sensor}
We implemented the quantum circuit shown in Fig.~\ref{fig:q_circuit}(d)
under standard NMR 
pulse sequences~\cite{liquidNMRQC}. 
The rotation operations applied to 
CC are 
\begin{eqnarray}
 R_{\rm C}(\phi,\theta) 
&=& R(\phi,\theta)\otimes \sigma_0 \otimes \sigma_0, \\
 Z_{\rm C}(\theta) 
&=& Z(\theta)\otimes \sigma_0 \otimes \sigma_0,
\end{eqnarray}
where 
$ R(\phi,\theta)= e^{-i \theta(\sigma_x \cos \phi 
+ \sigma_y\sin \phi)/2}$ 
and $Z(\theta) = e^{-i \theta \sigma_z/2}$. 
$\theta $ in $R(\theta,\phi)$ is the rotation angle and $\phi$ defines 
the rotation axis in the $xy$-plane from the $x$-axis, while $\theta$ in 
$Z(\theta)$ is the rotation angle about
the $z$-axis. 
The rotations that operate on the CSs
can be implemented simultaneously
because of the symmetry of the molecular structure and are defined as
\begin{eqnarray}
  R_{\rm S}(\phi,\theta) 
&=& \sigma_0 \otimes R(\phi,\theta)\otimes R(\phi,\theta), \\
 Z_{\rm S}(\theta) 
&=& \sigma_0 \otimes Z(\theta)\otimes  Z(\theta). 
\end{eqnarray}
The Hadamard gate on CC was effectively implemented 
with an $R_{\rm C}(\pi/2,-\pi/2)$, 
while a pseudo-CNOT gate was realized as follows:
\begin{eqnarray}
 {\rm CNOT} 
&=& e^{ -i \frac{\pi}{4}} Z_{\rm C}(-\frac{\pi}{2}) Z_{\rm S}(-\frac{\pi}{2}) 
R_{\rm C}(0, \frac{\pi}{2})  U_{\rm E} R_{\rm S}(\frac{\pi}{2}, \frac{\pi}{2}) 
\nonumber \\
&=&  
{\small  \left(
\begin{array}{cccccccc}
 1 & 0 & 0 & 0 & 0 & 0 & 0 & 0 \\
 0 & 1 & 0 & 0 & 0 & 0 & 0 & 0  \\
 0 & 0 & 1 & 0 & 0 & 0 & 0 & 0  \\
 0 & 0 & 0 & 1 & 0 & 0 & 0 & 0 \\
 0 & 0 & 0 & 0 & 0 & 0 & 0 & i \\
 0 & 0 & 0 & 0 & 0 & 0 & i & 0 \\
 0 & 0 & 0 & 0 & 0 & i & 0 & 0 \\
 0 & 0 & 0 & 0 & i & 0 & 0 & 0 \\
\end{array} \right) }, 
\end{eqnarray}
where 
$U_{\rm E} = e^{-\pi(\sigma_z \otimes \sigma_z \sigma_0 +
\sigma_z \otimes \sigma_0 \sigma_z)/4}$. 
$U_{\rm E}$ was implemented by waiting for a period of 
$\displaystyle \frac{n}{\Delta \omega_0}$, where $\Delta \omega_0$
is the Larmor frequency difference between CC and CSs.  
$n$ is an integer and is selected so that 
$ \displaystyle \frac{n}{\Delta \omega_0} 
\approx \frac{\pi}{J^{\rm (CC, CSs)}}$ (see Table~\ref{H_parameters}). 
All $Z_k(\theta)$'s ($k = $ C or S) 
were virtually implemented 
by controlling the $\phi$'s in the 
$R_k(\phi, \theta)$'s  ($k = $ C or S)~\cite{Kondo2006}.  
We employed 
jump-and-return pulses~\cite{PhysRevA.74.052324} to 
realize $R_k(\phi, \theta)$ 
with concatenated composite pulses 
(reduced CinBB)~\cite{doi:10.7566/JPSJ.82.014004,SR_Comp_Pulse_2020}
to reduce any errors caused by imperfect pulses. 

Magnetic field sensing is equivalent to measuring the 
phase difference between the initial and final state 
of CC, as shown in Fig.~\ref{fig:q_circuit}. Therefore,
we simulated the magnetic field by applying a $Z_k(\theta)$-rotation,
\begin{align*}
&\textbf{field on CC:}\\
&\left(R_{\rm C}(-\frac{\pi}{2}, \frac{\pi}{2})-Z_{\rm C}(\theta)\right)-
{\rm CNOT}-(\frac{\tau}{2}-R_{\rm S}(0,\pi)-\frac{\tau}{2})-{\rm CNOT}, \\
&\textbf{field on CSs:}\\
&R_{\rm C}(-\frac{\pi}{2},\frac{\pi}{2})-\left({\rm CNOT}-Z_{\rm S}(\theta)\right)-
(\frac{\tau}{2}-R_{\rm S}(0,\pi)-\frac{\tau}{2}) - {\rm CNOT}, 
\end{align*}
where $(\tau/2-R_{\rm S}(0,\pi)-\tau/2)$ is the period 
when the entangled 
sensor acquires the phase $\theta = \gamma_{\rm G} B \tau$ 
in real measurements.
In our simulations, the ``entangled magnetic sensor'' 
(CSs) is active in the 
controlled environment
only during this period. $R_{\rm S}(0, \pi)$ 
in the middle of 
this period was added to nullify the time development caused by the 
interaction between CC and CSs. 
We started from the thermal state and observed only CC   
in our simulations. 

\begin{figure}[t]
\begin{center}
\includegraphics[width=8.6cm]{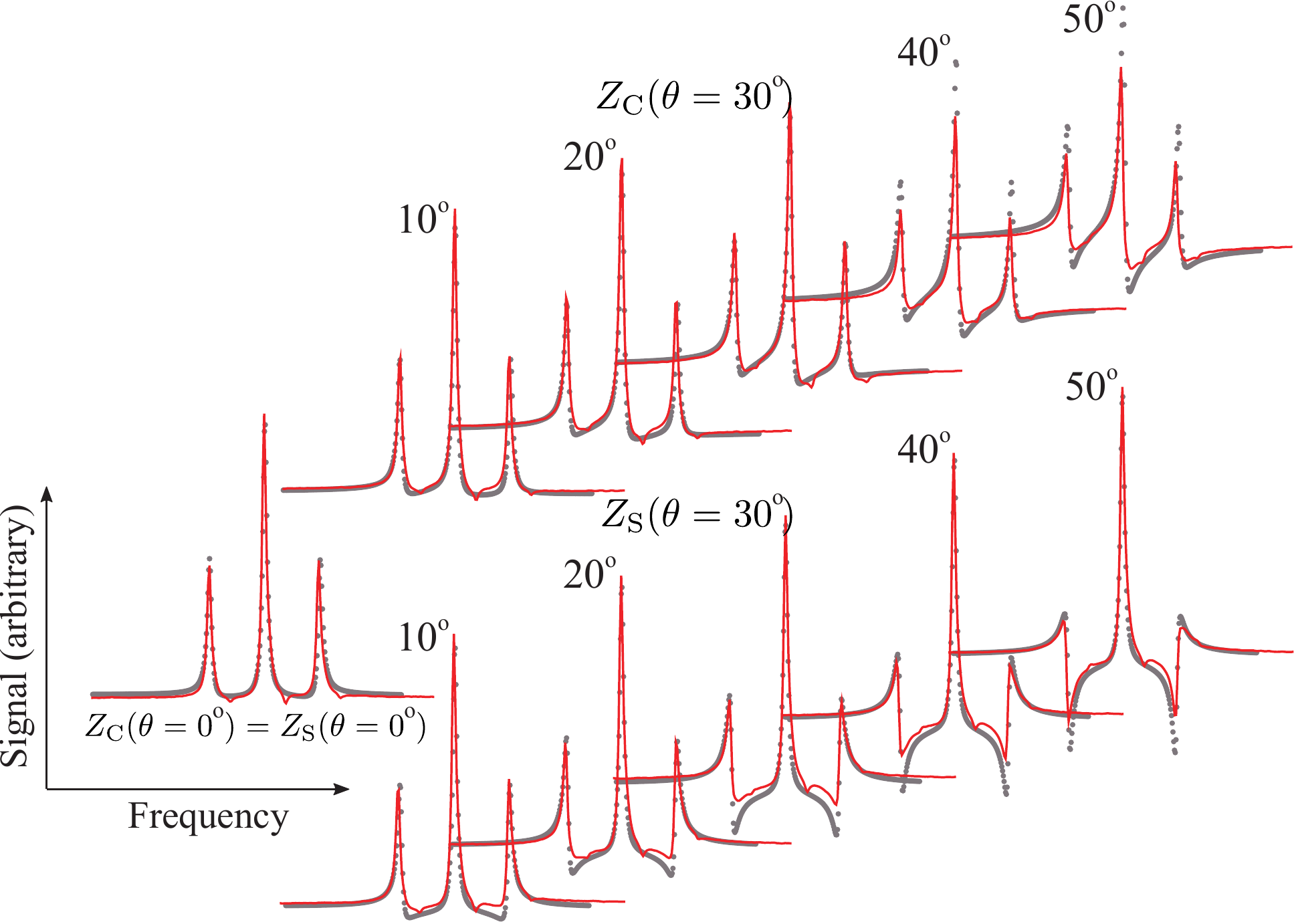}
\vspace{-0.5cm}
\caption{(Color online)
Spectra as functions of $\theta$ (the strength of the 
``magnetic field'')
when all H spins were decoupled (full-decoupling case). 
The (red) solid lines are measured spectra, while the grey dots 
are the calculated theoretical spectra, such as in Fig.~\ref{fig_fid_th}. 
The left-most spectra is the reference without a 
``magnetic field''. The upper spectra show cases when ``magnetic field''s were
applied to CC (non-entangled sensor), 
while the lower ones correspond to CSs (the entangled sensor). 
The frequencies of the center peaks are 62.6~ppm and 
the frequency differences among the peaks are $2\pi\cdot 38.4$~rad/s,
as listed in Table~\ref{H_parameters}.
}
\label{fig:p_meas}
\end{center}
\end{figure}

We first demonstrate the ``entanglement-enhanced''
phase sensitivity~\cite{Jones1166,Liu:2015aa,Knott:2016aa}
in the full-decoupling case (Fig.~\ref{fig_st}(c)).
A sample consisting of 0.41~M ${}^{13}$C-labeled
2-propanol solved in acetone-d6 with 
12~mM of Fe(acac) as a magnetic impurity was used. The $T_1$'s of 
$^{13}$C of this sample were measured to be 1.3~s, 
while the $T_1$'s of all H spins were approximately 100~ms.  
Figure~\ref{fig:p_meas} shows the spectra of CC  
as a function of $\theta$ (the ``field strength'') 
at $\tau = 3.4$~ms.  
When the ``field'' was applied to CC
(Fig. \ref{fig:p_meas} (upper), 
the non-entangled sensor), 
the three peaks 
acquired the same amount of phase. These phases were 
the same as $\theta$ given by $Z_{\rm C}(\theta)$
within an acceptable experimental error range. 
Note the good agreement between the measured 
and calculated spectra. On the other hand, 
the acquired phases were 
depending on peaks
when the ``field'' was applied to CSs
(Fig. \ref{fig:p_meas} (lower), entangled sensor). 
The center peak does not 
acquire any phase, as one can see from the fact that 
it is symmetric, regardless of the $\theta$ values. 
The two side peaks acquire $\pm 2 \theta $;
the $+$ sign is for the left peak and the $-$ sign is for the right. 
Again, we obtain reasonably good agreement between 
the measured and calculated spectra, 
although the measured spectra are not as sharp as the calculated spectra. 
The case of the non-entangled sensor appears to 
result in better agreement between the experiments and theory compared to 
that of the entangled sensor. We explain this result as follows. 
The phase (``field'') information is stored in the entangled spins ($\circ$'s) 
in the latter case and we therefore expect that this information is more 
fragile than that in the former case because of the fragility of the entangled state.
All calculated spectra were obtained with the measured coupling constant of
$J^{(\rm CC, CSs)}=2\pi \cdot 38.4$~rad/s and two fitting parameters: 
one being $T_2 = 0.1$~s and 
the other being the amplitude. The value of $T_2$ is quite reasonable 
based on the FID signal measurements. Therefore, the amplitude is 
the only fitting parameter required. 

\begin{figure}[t]
\begin{center}
\includegraphics[width=8.6cm]{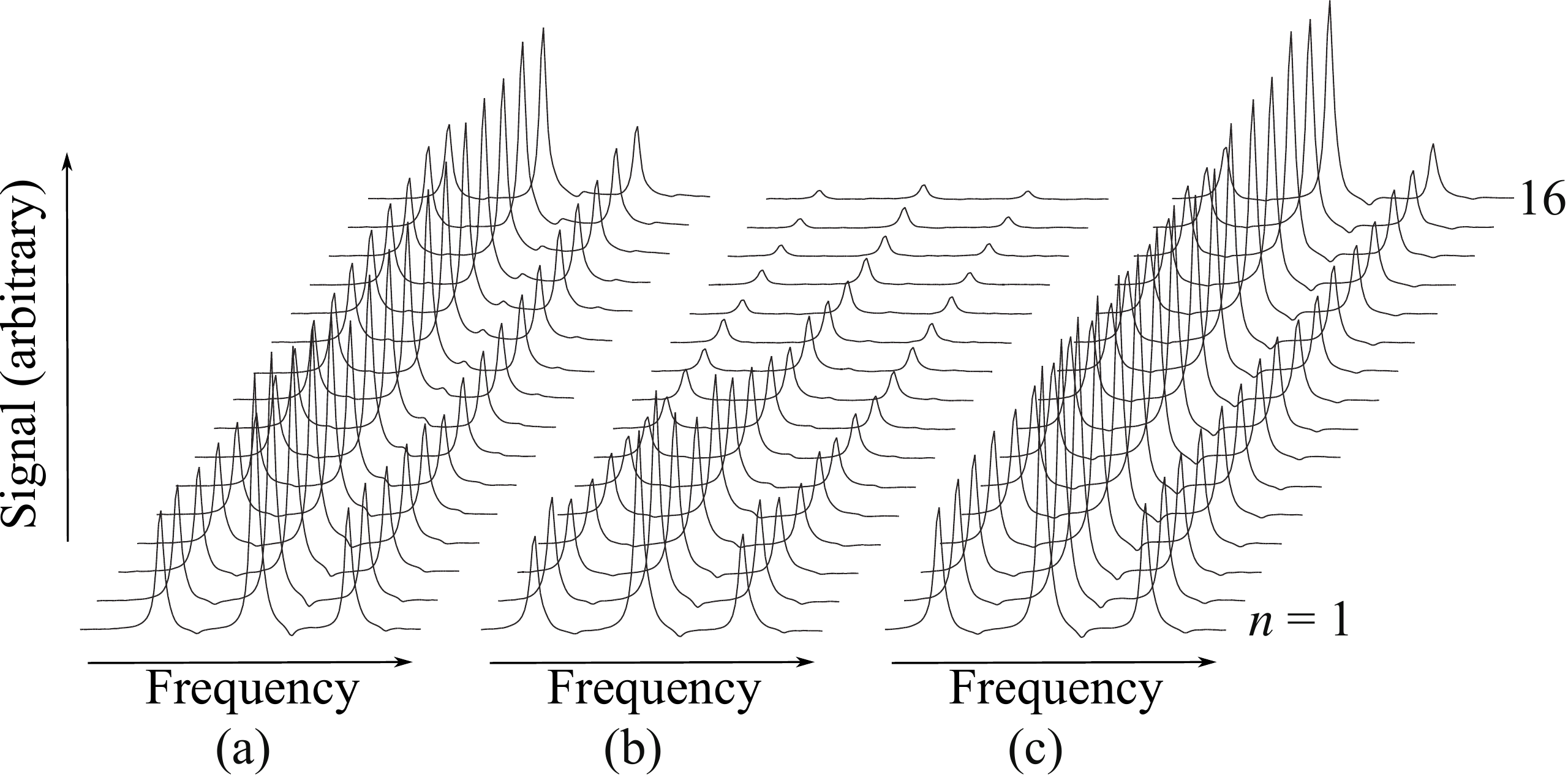}
\caption{
Measurements in various noisy environments
as a function of measurement time $\tau(=3.44~{\rm ms} \times n)$. 
(a) ``without noise'' by decoupling all H spins (full-decoupling case),
(b) ``under noise'' without decoupling HSs (selective-decoupling), and 
(c) ``under noise,'' where the noise is suppressed by a 
dynamical decoupling technique (XY-8) 
during measurements. 
The frequency of the center peaks is 62.6~ppm and 
the frequency differences among the peaks are $2\pi\cdot 38.4$~rad/s,
as listed in Table~\ref{H_parameters}.
}
\label{fig:sen_rel}
\end{center}
\end{figure}

Next, we show 
that the ``magnetic field sensing'' was affected by ``noise'' 
generated by the controlled environment
and that the ``noise'' 
can be suppressed 
by a dynamical decoupling technique 
(XY-8~\cite{RevModPhys.88.041001}). 
Here, HS (corresponding to a $\bullet$ in 
Fig.~\ref{fig:q_circuit}(c)) and magnetic impurities 
combine to generate a time-inhomogeneous noisy environment 
acting on the entangled sensor (that corresponds to the $\circ$'s in 
Fig.~\ref{fig:q_circuit}(c)). We can nullify 
the time-inhomogeneous noisy environment 
by effectively removing the HSs with a decoupling technique. 
The experimental details are 
as follows. 
We consider two cases: the full-decoupling case (Fig.~\ref{fig_st}(c)) 
and the selective-decoupling case (Fig.~\ref{fig_st}(b)). 
In the full-decoupling case, 
the magnetic impurities produce weak effects on CC and CSs~\cite{Binho2019}, 
and thus we can state that 
``magnetic field sensing'' was performed under an approximately noiseless enviroment in the short time scale of less than 
50~ms in the experiment. This 
approximation was confirmed by the fact that 
the signal exhibited little decay in this time scale
(see Fig.~\ref{fig:sen_rel}(a)).
In the selective decoupling case, CSs should be affected by
the time-inhomogeneous noisy environment formed by the 
HSs and the magnetic impurities. This noisy environment was also
confirmed by the fact that the signal decays quickly, as 
shown in Fig.~\ref{fig:sen_rel}(b). 
The relaxation constant is approximately 30~ms.
The spectra in Fig.~\ref{fig:sen_rel}(c) were measured when the 
HSs were not decoupled (the same as (b)), but the XY-8 
sequence was applied to CC and CSs simultaneously. 
The signals decay much more slowly than 
those in (b), which indicates that the dynamical decoupling was 
effective in protecting the entangled sensor. 
When a dynamical decoupling technique is applied 
to a sensor, it cannot detect the DC component, but can measure 
the AC one, whose frequency is determined by the decoupling 
technique~\cite{RevModPhys.89.035002}. 
Therefore, it is possible to construct
an entanglement-enhanced AC magnetic field sensor under time-inhomogeneous 
noise, as theoretically predicted in Refs.\ 
\cite{PhysRevA.84.012103,chin2012quantum,tanaka2015proposed,RevModPhys.88.021002}.


\section{Summary}\label{sec4}
We have successfully modeled an entangled magnetic field sensor 
in various noisy environments using NMR techniques. 
In our model, a sensor is a star-topology molecule, 2-propanol, 
solved in acetone-d6, and the magnetic field is simulated 
by rotational pulse sequences acting on the sensor. 
The environment surrounding a sensor can be controlled by 
adding Fe(III) as an impurity in the solvent and 
by selectively decoupling H spins in the 2-propanol molecule. 
We have demonstrated the entanglement-enhanced phase sensitivity and have discussed its enhancement mechanism. 
We further demonstrate that magnetic field sensing is affected 
by noise. Importantly, we have demonstrated that 
when the noise is time-inhomogeneous, 
its effect can be suppressed by a dynamical 
decoupling technique during the entanglement-enhanced magnetic field sensing period.

\section*{Acknowledgments}
This work was supported by CREST(JPMJCR1774), JST.
This work is also 
supported by Leading Initiative for 
Excellent Young Researchers MEXT Japan and
MEXT KAKENHI (Grant No. 15H05870).

\appendix
\section{}
In previous work~\cite{Binho2019}, 
we systematically studied the three cases 
when the environments consisted of 
1, 3, and 12 spins $+$ a Markovian 
environment generated by magnetic impurities. 
We add another case study here involving a 
6 spins $+$ Markovian environment case by 
using 2-propanol solved in acetone d6. 

The longitudinal relaxation times, $T_1$'s of the $^{13}$C spins 
of a 0.41~M {}$^{13}$C-labeled 2-propanol sample 
solved in acetone d-6 without 
magnetic impurities were measured to be 20~s (CC) and 8~s (CSs). 
Therefore, within a time scale much shorter than these $T_1$'s, 
the $^{13}$C chain in the 2-propanol molecules can be approximated 
as isolated systems. We added magnetic impurities (Fe(III) acac)
and prepared four samples, as listed in Table~\ref{FID_tc}. 

\begin{table}[h]
\caption{\label{FID_tc}
Measured $T_1$'s and $T_2$'s of CC, and $T_1$ of HSs are summarized. 
$C_{\rm m}$: concentration of 
the magnetic impurity (Fe(III)acac), 
$T_1$: longitudinal relaxation time constant of CC, 
$T_2^{(\rm f)}$: relaxation time constant of 
the signal in the full-decoupling case,  
$T_2^{(\rm s)}$: relaxation time constant 
of the signal in the selective-decoupling case, and 
$T_1$(HSs): longitudinal relaxation time constant of HSs' spins. 
}
\begin{center}
\small\addtolength{\tabcolsep}{-0.5pt}
\begin{tabular}[t]{|c|c|c|c|c|c|c|c|}
\hline
\multirow{2}{*}{Sample}  &$C_{\rm m}$ &$T_1$ & $T_1 \cdot C_{\rm m}$
&$T_2^{(\rm f)}$ 
& $T_2^{(\rm f)}\cdot C_{\rm m}$& $T_2^{(\rm s)}$& $T_1$(HSs)\\
&(mM)&(s)&(mM$\cdot$s)&(ms)&(M$\cdot$s)&(ms)&(ms) \\
\hline 
1 &12 & 1.3 & 15 & $3.0\times 10^2$ 
& $3.5 \times 10^{-3}$& $3.0 \times 10$ &  93\\
\hline 
2 &26 & 0.64 & 17 & $1.0 \times 10^2$ 
& $2.6 \times 10^{-3}$& $3.9\times 10$ &  43\\
\hline 
3 &47 & 0.36& 17& $9.9\times 10$
& $4.7 \times 10^{-3}$ & $3.8\times 10$ &  24\\
\hline
4 &94& 0.17& 16& $6.4\times 10$
& $6.0 \times 10^{-3}$& $3.9\times 10$ &  17\\
\hline 
\end{tabular}
\end{center}
\end{table}

In the full-decoupling case, a small but not negligible 
direct influence of the Markovian environment on CC should 
be observed. The $T_1$'s of CC in Table~\ref{FID_tc} are 
inversely proportional to the magnetic 
impurity concentration $C_{\rm m}$, which implies that in this case, $T_1$ 
is determined by the impurity concentration~\cite{Iwakura2017}. 
On the other hand, in the selective-decoupling case, 
the interaction between CC and the Markovian environment 
through the HSs (System~II) should be added, although it is 
expected to be small. Therefore, we obtain the 
controlled environment,
which consists of 6 spins + Markovian 
environment, which causes a non-exponential decay 
of CC~\cite{Binho2019}.
We note, however, that the large interactions between CC and CSs
($J^{\rm (CC, CSs)}=2\pi \cdot 34$~rad/s) compared with those between CC and the HSs 
($J^{\rm (CC, HSs)}=2\pi \cdot 4.4$~rad/s)
prevent direct observation of the subtle non-exponential dynamics. 

\begin{figure}[t]
\begin{center}
\includegraphics[width=8.6cm]{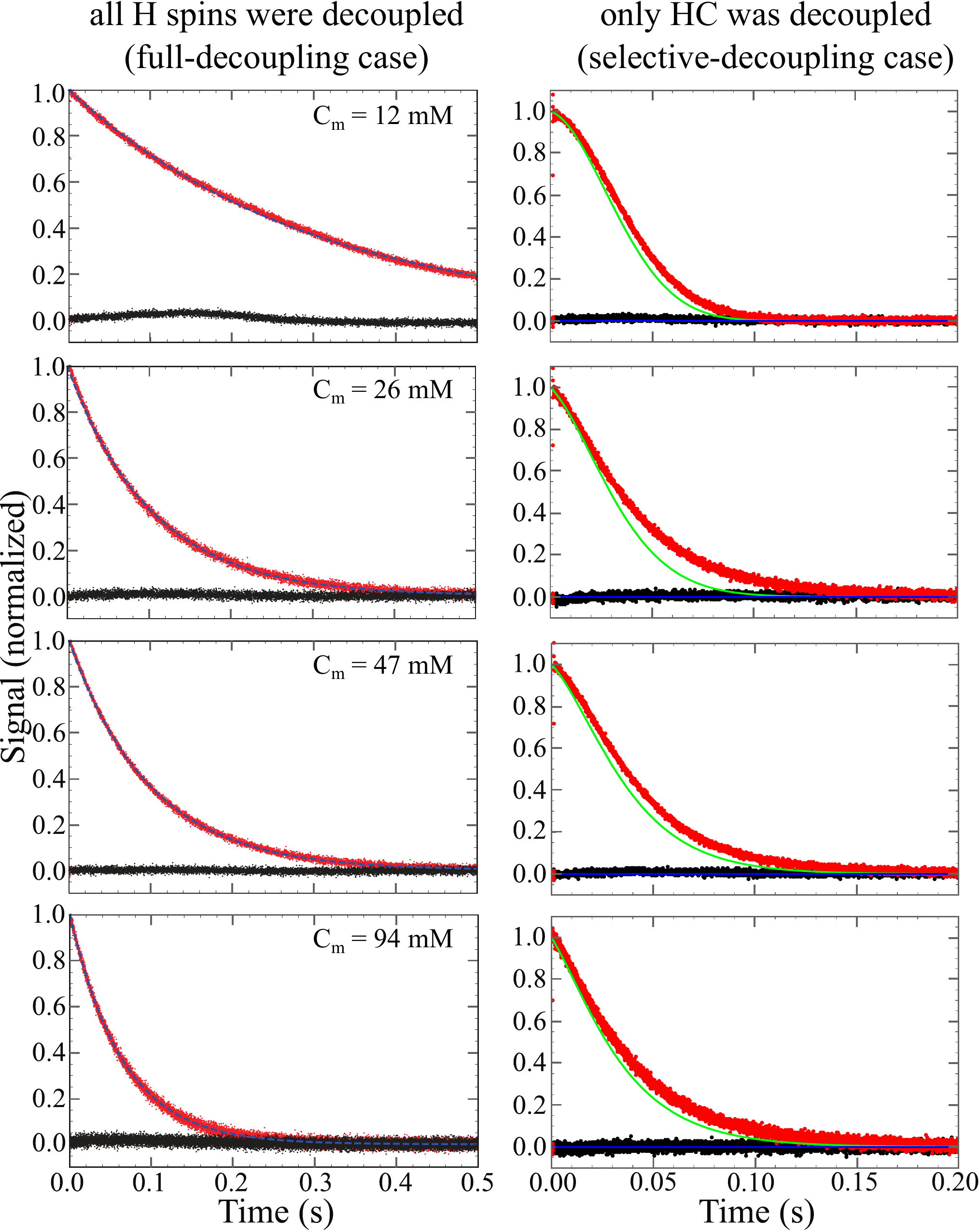}
\caption{(Color online)
FID signals of Samples~1, 2, 3, and 4 in Table~\ref{FID_tc}. 
The initial states were 
$\displaystyle |+\rangle \langle + | \otimes
\left( |01\rangle \langle 01 |
    +  |10\rangle \langle 10 | \right).
$ 
The real parts of the FID signals
are shown in red, while the imaginary are shown in black. 
The full-decoupling cases are shown in the left panels and 
selective decoupling cases are presented in the right panels. 
The black dashed curves in the left panels 
are exponential fittings to the real parts 
of the FID signals. 
The green (blue) curves in the right panels are the 
calculated real (imaginary) parts of the FID signals
\cite{Binho2019}.
The blue curves overlap the experimental data and 
thus are hardly visible.
}
\label{fig:fid}
\end{center}
\end{figure}

To successfully observe the above subtle non-exponential dynamics, 
let us re-examine the thermal state $\rho_{\rm th}$ of 
the three $^{13}$C's, which is~\cite{Cory1634,Gershenfeld350,liquidNMRQC}
\begin{eqnarray}
 \label{eq_thermal_state}
\rho_{\rm th}
&\approx & 
\underbrace{\frac{ \sigma_0 + \epsilon |0\rangle \langle 0 |}{2}}_{\rm CC}
\otimes \underbrace{\frac{\sigma_0}{2}\otimes\frac{\sigma_0}{2}}_{\rm CSs} 
\nonumber \\
&+&
\underbrace{\frac{ \sigma_0}{2}}_{\rm CC}
\otimes \underbrace{
\frac{ \sigma_0 + \epsilon |0\rangle \langle 0 |}{2}
\otimes\frac{\sigma_0}{2}}_{\rm CSs} \nonumber\\
&+&
\underbrace{\frac{ \sigma_0}{2}}_{\rm CC}
\otimes \underbrace{
\frac{\sigma_0}{2} \otimes 
\frac{ \sigma_0 + \epsilon |0\rangle \langle 0 |}{2}}_{\rm CSs}  
\end{eqnarray}
Here, $\epsilon 
\sim 10^{-5}$ in NMR measurements. 
When we observe only CC, $\rho_{\rm th}$ is equivalent to 
\begin{eqnarray*}
 \rho_{\rm th}
\approx
\frac{1}{8}
\left( \sigma_0 + \epsilon |0\rangle \langle 0 | \right)  
\otimes \left( |0\rangle\langle 0|+ |1\rangle\langle 1|\right)
\otimes \left( |0\rangle\langle 0|+ |1\rangle\langle 1|\right).
\end{eqnarray*}
Moreover, $\sigma_0$ of CC is not observable in NMR and thus 
$\rho_{\rm th}$ can be re-normalized as
\begin{eqnarray*}
 \rho_{\rm th}
\approx
\frac{1}{8}
\left( |0\rangle \langle 0 | \right)  
\otimes \left( |00\rangle\langle 00| + |01\rangle\langle 01|
+|10\rangle \langle 10| + |11\rangle \langle 11| \right).
\end{eqnarray*}
The interaction effects on CC from the $ |01\rangle\langle 01|$ and 
$|10\rangle \langle 10|$ states of CSs cancel each other out.
Thus
if we can prepare 
\begin{eqnarray}
\label{eq_rho_i}
 \rho_{\rm i} &=&  |+\rangle \langle + | \otimes 
\left( 
|01\rangle\langle 01|+|10\rangle \langle 10| 
\right),
\end{eqnarray}
we can observe the subtle non-exponential dynamics 
discussed previously. 
This $\rho_{\rm i}$
can be prepared with a standard NMR technique called a soft 
pulse~\cite{Levitt2008}. 

The results are summarized in Fig.~\ref{fig:fid}. 
We can successfully observe the exponential decays in the 
full decoupling cases (left panels), while the non-exponential decay
dynamics are observed in the selective decoupling cases (right panels).
We also calculated the decay dynamics from the data 
(summarized in Table~\ref{FID_tc}) as in our previous 
work~\cite{Binho2019}, which are plotted as green (blue) solid curves 
in the right panels in Fig.~\ref{fig:fid}. By taking into account 
that there are no fitting parameters except for the amplitude, 
we believe that the calculated dynamics reproduce the observations relatively well. 
However, the reproducibility may not be as good as in our preceding 
work~\cite{Binho2019}, which may be caused by the imperfect 
soft pulses employed for preparing the initial state or by 
the error in parameter determination summarized in Table~\ref{FID_tc}, 
especially regarding the interaction strength between CC and the HSs 
($J^{\rm (CC, HSs)}$).


\bibliography{bib18Oct}

\end{document}